\documentclass[journal,12pt,onecolumn,draftclsnofoot]{IEEEtran}

\usepackage{color}
\usepackage{multirow}
\usepackage{graphicx}
\usepackage{epstopdf}
\usepackage[cmex10]{amsmath}
\usepackage{amssymb}

\usepackage{multirow}
\usepackage{amsthm}
\usepackage{cite}
\usepackage{acronym} 
\usepackage{algorithm, algcompatible}

\newcommand{\diag}{\mathop{\rm diag}}

\algnewcommand\INPUT{\item[\textbf{Input:}]}
\algnewcommand\OUTPUT{\item[\textbf{Output:}]}

\allowdisplaybreaks

\begin{document}

\title{Semi-Blind Multiuser Detection Under the Presence of Reconfigurable Intelligent Surfaces} 
\author{Nikolaos~I.~Miridakis,~\IEEEmembership{Senior Member,~IEEE}, Theodoros~A.~Tsiftsis,~\IEEEmembership{Senior Member,~IEEE}, Guanghua~Yang,~\IEEEmembership{Senior Member,~IEEE}, Panagiotis~A.~Karkazis, and Helen~C.~Leligou
\thanks{\textit{Corresponding Author: T. A. Tsiftsis.}}
\thanks{N.~I.~Miridakis, T.~A.~Tsiftsis and G.~Yang are with the Institute of Physical Internet and School of Intelligent Systems Science \& Engineering, Jinan University, Zhuhai Campus, Zhuhai 519070, China. N.~I.~Miridakis is also with the Department of Informatics and Computer Engineering, University of West Attica, Aegaleo 12243, Greece (e-mails: nikozm@uniwa.gr, theo\_tsiftsis@jnu.edu.cn, ghyang@jnu.edu.cn).}
\thanks{P.~A.~Karkazis is with the Department of Informatics and Computer Engineering, University of West Attica, Aegaleo 12243, Greece (email: p.karkazis@uniwa.gr).}
\thanks{H.~C.~Leligou is with the Department of Industrial Design and Production Engineering, University of West Attica, Aegaleo 12241, Greece (email: e.leligkou@uniwa.gr).}
}


\maketitle

\begin{abstract}
A multiuser multiple-input multiple-output wireless communication system is analytically studied, which operates with the aid of a reconfigurable intelligent surface (RIS). The intermediate RIS is equipped with multiple elements and operates via random phase rotations to simultaneously serve multiple users. Independent Rayleigh fading conditions are assumed among the included channels. The system performance is analytically studied when the linear yet efficient zero-forcing detection is implemented at the receiver. In particular, the outage performance is derived in closed-form expression for different system configuration setups with regards to the available channel state information at the receiver. Further, a joint coherent/noncoherent linear detection is analytically presented. Finally, some new engineering insights are provided, such as how the channel state information and/or the volume of antenna/RIS arrays impact on the overall system performance as well as the arising efficiency on the performance/complexity tradeoff by utilizing the joint coherent/noncoherent scheme. 
\end{abstract}


\IEEEpeerreviewmaketitle

\section{Introduction}
\IEEEPARstart{T}{he} reconfigurable intelligent surfaces (RIS), also known as intelligent reflecting metasurfaces, are emerging at the forefront of wireless communications research nowadays \cite{j:WangLiu2020,j:QianRenzo21,j:ZhangHuahua21,j:PsomasKrikidis2021}. RISs can control the wireless channel by suitably shifting the phase (and the amplitude response) of the impinging signal on a large number of meta-elements. Doing so, the corresponding reflected signals can be added either constructively or destructively with other signals to enhance the signal-to-noise ratio (SNR) and/or to suppress the co-channel interference at the receiver. Most of the current art is focused on the incident signals' phase shift optimization and/or knowledge acquisition of the channel state information (CSI) at RIS \cite{j:DharmawansaAtapattu2021}. Nevertheless, such a condition reflects on a considerably high computational complexity and power consumption. In addition, assuming perfect CSI at RIS is challenging, while it may also be infeasible in some practical cases, due to the limited resources and passive nature of RIS deployments. The latter constraints become even more emphatic in multiuser multiple input-multiple output (MIMO) systems \cite{j:Khaleel2020}.

Motivated by the aforementioned observations, in this letter, the performance of a multiuser MIMO communication system is analyzed, which operates with the aid of an intermediate RIS. The linear zero-forcing (ZF) detection is utilized at the receiver to recover the simultaneously detected multiple signals. It is assumed that the signals undergo independent Rayleigh fading conditions. The practical scenario of a fully-blind RIS (in terms of the multiuser CSI) is considered, whereas it is assumed that the phases of RIS elements are randomly rotated so as to serve a set of arbitrarily located multiple users. 

New closed-form expressions with regards to the system outage probability are presented, under different use cases regarding the availability of instantaneous CSI at the receiver side. Moreover, the performance of a joint coherent/noncoherent detection scheme is analytically studied, which shows a remarkable efficiency on the performance/complexity tradeoff at RIS-enabled system setups. Insightfully, it is shown that the presence of partial-only CSI (which occurs whenever the CSI of at least one of the three links between transmitter, RIS and receiver is missing) unavoidably produces an SNR-independent outage/error floor. On the other hand, despite the partial CSI assumption, the introduced joint coherent/noncoherent detection effectively tackles this problem even for the suboptimal yet practically feasible random phase rotation strategy implemented at RIS. Also, the beneficial role of an increased antenna/RIS element array is manifested, while some useful engineering insights are presented. Most importantly, it turns out that utilizing the said joint detection scheme, the arising channel gains reflect on quite a similar performance as the ideal detection scheme (with full CSI of all the included links, yet impractical), but preserving much less computational costs and signaling overhead at the same time. 

{\it Notation:} Vectors and matrices are represented by lowercase and uppercase bold typeface letters, respectively, whereas $\mathbf{I}_{v}$ stands for the $v\times v$ identity matrix. A diagonal matrix with entries $x_{1},\cdots,x_{n}$ is defined as $\diag\{x_{i}\}^{n}_{i=1}$, $\mathbf{X}^{-1}$ is the inverse of $\mathbf{X}$ and $\mathbf{X}^{\dagger}$ is the Moore-Penrose pseudoinverse of $\mathbf{X}$. $[\mathbf{X}]_{k,l}$ stands for the entry at the $k^{\rm th}$ row and $l^{\rm th}$ column of $\mathbf{X}$. Superscript $(\cdot)^{\mathcal{H}}$ denotes Hermitian transposition, $|\cdot|$ represents absolute (scalar) value and $j\triangleq \sqrt{-1}$. $\mathbb{E}[\cdot]$ is the expectation operator, symbol $\overset{\text{d}}=$ means equivalence in distribution and $\overset{\text{d}}\asymp$ defines asymptotic equivalence in distribution. $f_{X}(\cdot)$ and $F_{X}(\cdot)$ represent the probability density function (PDF) and cumulative distribution function (CDF) of a random variable (RV) $X$, respectively. Furthermore, $F_{X|Y}(\cdot|\cdot)$ represents the distribution function of $X$ conditioned on $Y$. $\mathcal{CN}(\mu,\sigma^{2})$ defines a circularly symmetric complex-valued Gaussian RV with mean $\mu$ and variance $\sigma^{2}$; $\mathcal{X}^{2}_{v}$ stands for a (central) chi-squared RV with $v$ degrees-of-freedom (DoF); while $\mathcal{X}^{2}_{v}(u,w)$ denotes that $X$ is a non-central chi-squared RV with $v$ DoF, a non-centrality parameter $u$ and variance $w$. Also, $\Gamma(\cdot,\cdot)$ denotes the upper incomplete Gamma function \cite[Eq. (8.350.2)]{tables} and $\Gamma(\cdot)$ denotes the Gamma function \cite[Eq. (8.310.1)]{tables}. $Q_{1}(\cdot,\cdot)$ is the first-order Marcum-$Q$ function; ${}_1F_{1}(\cdot,\cdot;\cdot)$ is the Kummer's confluent hypergeometric function \cite[Eq. (9.210.1)]{tables}; and $G[\cdot|\cdot]$ represents the Meijer's $G$-function \cite[Eq. (9.301)]{tables}.

\section{System and Signal Model}
Consider a wireless communication system with $M$ single-antenna transmitters and a receiver equipped with $N\geq M$ antennas\footnote{From the analysis presented hereafter, the classical single-user communication scenario with $M$ co-located transmit antennas is included as a special case.} operating over a quasi-static block-fading channel. In fact, independent Rayleigh faded channels are assumed which remain fixed for the duration of a given frame transmission while they may change independently amongst various frames. The end-to-end communication is assisted by an RIS equipped with $L$ passive elements. It is also assumed that both the size of each element and the inter-element spacing are equal to half of the signal wavelength; such that the associated channels undergo independent fading \cite{j:RenzoZappone20}. All the given streams are simultaneously transmitted and received via ZF detection. 

More specifically, the received signal reads as 
\begin{align}
\mathbf{y}=\sqrt{p}\left(\mathbf{H}_{\rm D}+\mathbf{H}\mathbf{\Phi}\mathbf{G}\right)\mathbf{s}+\mathbf{n},
\label{received}
\end{align}
where $\mathbf{y} \in \mathbb{C}^{N \times 1}$, $p$ is the transmit SNR and $\mathbf{s}\in \mathbb{C}^{M \times 1}$ is the transmit signal. Also, $\mathbf{H}_{\rm D}\in \mathbb{C}^{N \times M}$ denotes the channel matrix between the transceiver direct links; $\mathbf{H}\in \mathbb{C}^{N \times L}$ and $\mathbf{G}\in \mathbb{C}^{L \times M}$ are the channel matrices of the RIS-to-receiver and transmitters-to-RIS links, respectively; $\mathbf{\Phi}\triangleq \diag\{e^{j \phi_{i}}\}^{L}_{i=1}$ denotes the phase rotations at RIS; and $\mathbf{n}\in \mathbb{C}^{N \times 1}$ defines the additive white Gaussian noise at the receiver. Further, it holds that $\mathbf{H}_{\rm D}\triangleq \mathbf{X}_{\rm D} \mathbf{\Xi}^{1/2}_{\rm D}$, where $\mathbf{X}_{\rm D}\in \mathbb{C}^{N \times M}$ and $\mathbf{X}_{\rm D}\overset{\text{d}}=\mathcal{CN}(\mathbf{0},\mathbf{I}_{N})$, while $\mathbf{\Xi}_{\rm D}\triangleq \diag\{\xi^{2}_{{\rm D},i}\}^{M}_{i=1}$ with $\xi^{2}_{{\rm D},i}$ standing for the (known) large-scale channel gain of the $i^{\rm th}$ transmitter. In a similar basis, $\mathbf{H}\triangleq \mathbf{X}_{\rm H} \mathbf{\Xi}^{1/2}_{\rm H}$, $\mathbf{X}_{\rm H}\in \mathbb{C}^{N \times L}$, $\mathbf{X}_{\rm H}\overset{\text{d}}=\mathcal{CN}(\mathbf{0},\mathbf{I}_{N})$, and $\mathbf{\Xi}_{\rm H}\triangleq \xi^{2}_{\rm H} \mathbf{I}_{L}$. Also, $\mathbf{G}\triangleq \mathbf{X}_{\rm G} \mathbf{\Xi}^{1/2}_{\rm G}$ with $\mathbf{X}_{\rm G}\in \mathbb{C}^{L \times M}$, $\mathbf{X}_{\rm G}\overset{\text{d}}=\mathcal{CN}(\mathbf{0},\mathbf{I}_{L})$, and $\mathbf{\Xi}_{\rm G}\triangleq \diag\{\xi^{2}_{{\rm G},i}\}^{M}_{i=1}$. The gains $\{\xi^{2}_{{\rm D},i},\xi^{2}_{{\rm G},i},\xi^{2}_{\rm H}\}$ incorporate signal propagation attenuation, antenna gains and shadowing losses, which are assumed either fixed or perfectly known at the receiver.\footnote{Note that the large-scale RIS-to-receiver channel gains $\xi^{2}_{\rm H}$ are identical for all the involved channel links since the inter-element distance at RIS (or the inter-antenna distance at the receiver) is considered negligible compared to the corresponding RIS-to-receiver distance.} Lastly, it is assumed that $\mathbb{E}[\mathbf{s}\mathbf{s}^{\mathcal{H}}]=\mathbf{I}_{M}$ and $\mathbf{n}\overset{\text{d}}=\mathcal{CN}(\mathbf{0},\mathbf{I}_{N})$.

Upon the signal reception, the ZF filter matrix $\mathbf{W}\in \mathbb{C}^{M \times N}$ is applied, yielding to the post-processed signal $\mathbf{r}=\mathbf{W}\mathbf{y}$. The structure of $\mathbf{W}$ depends on the implicit assumption of the known (instantaneous small-scale) channel gains and is further analyzed subsequently.

\section{System Performance}
In current work, we study the outage performance of the considered system configuration for different ZF filters $\mathbf{W}$. For a given SNR of the $i^{\rm th}$ transmitted stream, defined as $\gamma_{i}$, its outage probability is computed as $P_{\gamma_{i}}(\gamma_{\rm th})\triangleq F_{\gamma_{i}}(\gamma_{\rm th})$, where $\gamma_{\rm th}\triangleq 2^{R}-1$ is the outage threshold satisfying a target data transmission rate $R$ (in bps/Hz).

\subsection{Given $\mathbf{H}_{\rm D}$, $\mathbf{W}\triangleq \mathbf{H}^{\dagger}_{\rm D}$}
In this case, the receiver perfectly knows the channel condition of the direct transceiver links, while the individual RIS-enabled channels $\mathbf{H}$ and $\mathbf{G}$ as well as the phase rotations $\mathbf{\Phi}$ are completely unknown. It holds that
\begin{align}
\nonumber
\mathbf{r}&=\mathbf{H}^{\dagger}_{\rm D}\left[\sqrt{p}\left(\mathbf{H}_{\rm D}+\mathbf{H}\mathbf{\Phi}\mathbf{G}\right)\mathbf{s}+\mathbf{n}\right]\\
&=\sqrt{p} \mathbf{s}+\underbrace{\sqrt{p} \mathbf{H}^{\dagger}_{\rm D}\mathbf{H}\mathbf{\Phi}\mathbf{G}\mathbf{s}+\mathbf{H}^{\dagger}_{\rm D}\mathbf{n}}_{\mathbf{n}_{1}'},
\label{receivedCSIdirect}
\end{align}
where $\mathbf{n}_{1}'$ denotes the post-processed noise vector with a corresponding covariance matrix being
\begin{align}
\nonumber
&\mathbb{E}\left[\mathbf{n}_{1}'(\mathbf{n}_{1}')^{\mathcal{H}}\right]\\
\nonumber
&=p \mathbf{H}^{\dagger}_{\rm D}\mathbb{E}\left[\mathbf{H}\mathbf{\Phi}\mathbf{G}\mathbf{G}^{\mathcal{H}}\mathbf{\Phi}^{\mathcal{H}}\mathbf{H}^{\mathcal{H}}\right]\left(\mathbf{H}^{\dagger}_{\rm D}\right)^{\mathcal{H}}+\mathbf{H}^{\dagger}_{\rm D}\left(\mathbf{H}^{\dagger}_{\rm D}\right)^{\mathcal{H}}\\
\nonumber
&=\mathbf{H}^{\dagger}_{\rm D}\left(\mathbf{H}^{\dagger}_{\rm D}\right)^{\mathcal{H}}\left(p L \xi^{2}_{\rm H} \sum^{M}_{i=1}\xi^{2}_{{\rm G},i}\right)+\mathbf{H}^{\dagger}_{\rm D}\left(\mathbf{H}^{\dagger}_{\rm D}\right)^{\mathcal{H}}\\
&=\left(\mathbf{H}^{\mathcal{H}}_{\rm D}\mathbf{H}_{\rm D}\right)^{-1}\left(p L \xi^{2}_{\rm H} \sum^{M}_{i=1}\xi^{2}_{{\rm G},i}+1\right),
\label{covDirect}
\end{align}
where we used the fact that $\mathbf{H}$ and $\mathbf{G}$ are mutually independent; $\mathbb{E}\left[\mathbf{H}\mathbf{H}^{\mathcal{H}}\right]=L \xi^{2}_{\rm H} \mathbf{I}_{N}$; $\mathbf{\Phi} \mathbf{\Phi}^{\mathcal{H}}=\mathbf{I}_{L}$; $\mathbb{E}\left[\mathbf{G}\mathbf{G}^{\mathcal{H}}\right]=\sum^{M}_{i=1}\xi^{2}_{{\rm G},i} \mathbf{I}_{L}$; and $\mathbf{H}^{\dagger}_{\rm D}(\mathbf{H}^{\dagger}_{\rm D})^{\mathcal{H}}=(\mathbf{H}^{\mathcal{H}}_{\rm D}\mathbf{H}_{\rm D})^{-1}$. Thereby, the received SNR of the $i^{\rm th}$ stream becomes
\begin{align}
\gamma_{i}=\frac{p}{\left(p L \xi^{2}_{\rm H} \sum^{M}_{i=1}\xi^{2}_{{\rm G},i}+1\right)\left[\left(\mathbf{H}^{\mathcal{H}}_{\rm D}\mathbf{H}_{\rm D}\right)^{-1}\right]_{i,i}}.
\label{snrDirect}
\end{align}
Obviously, even when operating at the high transmit SNR regime, the received SNR does not infinitely grow without any bound, since $\gamma_{p\rightarrow \infty}=\{(L \xi^{2}_{\rm H} \sum^{M}_{i=1}\xi^{2}_{{\rm G},i})[(\mathbf{H}^{\mathcal{H}}_{\rm D}\mathbf{H}_{\rm D})^{-1}]_{i,i}\}^{-1}$, thereby, introducing a certain outage/error floor. It is well-known that $\gamma$, as per \eqref{snrDirect}, follows a scaled $\chi^{2}_{N-M+1}$ distribution with a corresponding outage probability given by \cite{j:GOREHeathPaulraj}
\begin{align}
P_{\gamma_{i}}(\gamma_{\rm th})=1-\frac{\Gamma\left(N-M+1,\frac{\gamma_{\rm th}\left(p L \xi^{2}_{\rm H} \sum^{M}_{i=1}\xi^{2}_{{\rm G},i}+1\right)}{p \xi^{2}_{{\rm D},i}}\right)}{\Gamma(N-M+1)}.
\label{poutDirect}
\end{align}

\subsection{Given $\mathbf{H}\mathbf{\Phi}\mathbf{G}$, $\mathbf{W}\triangleq \left(\mathbf{H}\mathbf{\Phi}\mathbf{G}\right)^{\dagger}$}
In this case, the channel gains between the transmitter-RIS-receiver link are perfectly known at the receiver\footnote{As an indicative example, the transmitter-RIS-receiver cascaded channel can be evaluated by using conventional channel estimation techniques, such as the linear minimum mean squared error or deep learning approach \cite{j:KunduMcKay21}.} (not at RIS), whereas the relevant channel status of the direct link remains unknown (or is absent). Following quite a similar approach as in the previous subsection, we get
\begin{align}
\nonumber
\mathbf{r}&=\left(\mathbf{H}\mathbf{\Phi}\mathbf{G}\right)^{\dagger}\left[\sqrt{p}\left(\mathbf{H}_{\rm D}+\mathbf{H}\mathbf{\Phi}\mathbf{G}\right)\mathbf{s}+\mathbf{n}\right]\\
&=\sqrt{p} \mathbf{s}+\underbrace{\sqrt{p} \left(\mathbf{H}\mathbf{\Phi}\mathbf{G}\right)^{\dagger}\mathbf{H}_{\rm D}\mathbf{s}+\left(\mathbf{H}\mathbf{\Phi}\mathbf{G}\right)^{\dagger}\mathbf{n}}_{\mathbf{n}_{2}'},
\label{receivedCSIRIS}
\end{align}
and
\begin{align}
\nonumber
&\mathbb{E}\left[\mathbf{n}_{2}'(\mathbf{n}_{2}')^{\mathcal{H}}\right]\\
\nonumber
&=p \left(\mathbf{H}\mathbf{G}\right)^{\dagger}\mathbb{E}\left[\mathbf{H}_{\rm D}\mathbf{H}^{\mathcal{H}}_{\rm D}\right]\left[\left(\mathbf{H}\mathbf{G}\right)^{\dagger}\right]^{\mathcal{H}}+\left(\mathbf{H}\mathbf{G}\right)^{\dagger}\left[\left(\mathbf{H}\mathbf{G}\right)^{\dagger}\right]^{\mathcal{H}}\\
\nonumber
&=\left(\mathbf{H}\mathbf{G}\right)^{\dagger}\left[\left(\mathbf{H}\mathbf{G}\right)^{\dagger}\right]^{\mathcal{H}}\left(p \sum^{M}_{i=1}\xi^{2}_{{\rm D},i}\right)+\left(\mathbf{H}\mathbf{G}\right)^{\dagger}\left[\left(\mathbf{H}\mathbf{G}\right)^{\dagger}\right]^{\mathcal{H}}\\
&=\left[\left(\mathbf{H}\mathbf{G}\right)^{\mathcal{H}}\left(\mathbf{H}\mathbf{G}\right)\right]^{-1}\left(p \sum^{M}_{i=1}\xi^{2}_{{\rm D},i}+1\right).
\label{covRIS}
\end{align}
Notably, $\mathbf{\Phi}$ is not present at the noise covariance of \eqref{covRIS} since, in principle, random phase rotations and reflections do not affect channel gains. Hence, the received SNR yields as
\begin{align}
\gamma_{i}=\frac{p}{\left(p \sum^{M}_{i=1}\xi^{2}_{{\rm D},i}+1\right)\left[\left[\left(\mathbf{H}\mathbf{G}\right)^{\mathcal{H}}\left(\mathbf{H}\mathbf{G}\right)\right]^{-1}\right]_{i,i}}.
\label{snrRIS}
\end{align}
Similar to the previous case, the latter SNR is bounded even in the asymptotically high $p$ regions; that is $\gamma_{p\rightarrow \infty}=\{(\sum^{M}_{i=1}\xi^{2}_{{\rm D},i})[[(\mathbf{H}\mathbf{G})^{\mathcal{H}}(\mathbf{H}\mathbf{G})]^{-1}]_{i,i}\}^{-1}$. The received SNR as per \eqref{snrRIS} can be considered as the product of two independent chi-squared RVs with $N-L+1$ and $L-M+1$ degrees-of-freedom, correspondingly. Doing so, its PDF reads as \cite[Eq. (26)]{c:Caire19}
\begin{align}
\nonumber
f_{\gamma_{i}}(x)&=\frac{\left(p \sum^{M}_{i=1}\xi^{2}_{{\rm D},i}+1\right)}{(N-M)!(L-M)!p \xi^{2}_{\rm H} \xi^{2}_{{\rm G},i}}\\
&\times G^{2,0}_{0,2}\left[\frac{\left(p \sum^{M}_{i=1}\xi^{2}_{{\rm D},i}+1\right) x}{p \xi^{2}_{\rm H} \xi^{2}_{{\rm G},i}}~\vline
\begin{array}{c}
\scriptstyle - \\
\scriptstyle N-M,L-M
\end{array}\right].
\label{PDFsnrRIS}
\end{align}
Then, utilizing \cite[Eq. (2.24.2.2)]{b:prudnikovvol3}, the outage probability arises as
\begin{align}
\nonumber
P_{\gamma_{i}}(\gamma_{\rm th})&=\frac{\left(p \sum^{M}_{i=1}\xi^{2}_{{\rm D},i}+1\right) \gamma_{\rm th}}{(N-M)!(L-M)!p \xi^{2}_{\rm H} \xi^{2}_{{\rm G},i}}\\
&\times G^{2,1}_{1,3}\left[\frac{\left(p \sum^{M}_{i=1}\xi^{2}_{{\rm D},i}+1\right) \gamma_{\rm th}}{p \xi^{2}_{\rm H} \xi^{2}_{{\rm G},i}}~\vline
\begin{array}{c}
\scriptstyle 0 \\
\scriptstyle N-M,L-M,-1
\end{array}\right].
\label{poutRIS}
\end{align}

\subsection{Given $\mathbf{H}_{\rm D}+\mathbf{H}\mathbf{\Phi}\mathbf{G}$, $\mathbf{W}\triangleq \left(\mathbf{H}_{\rm D}+\mathbf{H}\mathbf{\Phi}\mathbf{G}\right)^{\dagger}$}
This scenario reflects on the ideal condition when the receiver perfectly knows the status of all the included channel fading links, which may serve as a performance benchmark. Unfortunately, the explicit formation of $\mathbf{W}$ renders the problem not amenable for further analysis in terms of an exact and closed-form outage expression. Nonetheless, keeping in mind that $L\gg 1$ in the vast majority of RIS deployments (typically, the number of passive reflecting elements is moderately high in realistic scenarios) and adopting the Lyapunov central limit theorem, we can easily show that\footnote{In principle, it is required that $L\rightarrow \infty$. However, it has been shown in the literature (e.g., see \cite{j:Badiu2020}) that \eqref{CLTapprox} still holds even for moderate $L$ ranges (say, $L\geq 16$) . This will be numerically verified in the next Section.} 
\begin{align}
\mathbf{H}\mathbf{\Phi}\mathbf{G}\overset{\text{d}}\asymp \mathbf{A}\mathbf{\Psi}^{1/2}, \textrm{ as } L\rightarrow \infty, 
\label{CLTapprox}
\end{align}
where $\mathbf{A}\overset{\text{d}}=\mathcal{CN}(\mathbf{0},\mathbf{I}_{N})$ and $\mathbf{\Psi}=\diag\{\xi^{2}_{\rm H}\xi^{2}_{{\rm G},i}/L\}^{M}_{i=1}$. Therefore, we proceed by letting the ZF filter be $\mathbf{W}\triangleq \left(\mathbf{H}_{\rm D}+\mathbf{H}\mathbf{\Phi}\mathbf{G}\right)^{\dagger}\approx \left(\mathbf{H}_{\rm D}+\mathbf{A}\mathbf{\Psi}^{1/2}\right)^{\dagger}$, while the detected signal is approached by
\begin{align}
\mathbf{r}\approx \sqrt{p} \mathbf{s}+\underbrace{\left(\mathbf{H}_{\rm D}+\mathbf{A}\mathbf{\Psi}^{1/2}\right)^{\dagger}\mathbf{n}}_{\mathbf{n}_{3}'}.
\label{approx}
\end{align}
Hence, the received SNR in this case yields as
\begin{align}
\gamma_{i}\approx \frac{p}{\left\{\left[\left(\mathbf{H}_{\rm D}+\mathbf{A}\mathbf{\Psi}^{1/2}\right)^{\mathcal{H}}\left(\mathbf{H}_{\rm D}+\mathbf{A}\mathbf{\Psi}^{1/2}\right)\right]^{-1}\right\}_{i,i}}.
\label{snrapprox}
\end{align}
Since both $\mathbf{H}_{\rm D}$ and $\mathbf{A}$ are mutually independent with zero-mean circularly symmetric complex-Gaussian entries while $\mathbf{\Psi}$ has a diagonal structure, the corresponding outage probability has a well-known form and is directly expressed as
\begin{align}
P_{\gamma_{i}}(\gamma_{\rm th})\approx 1-\frac{\Gamma\left(N-M+1,\frac{\gamma_{\rm th}}{p \left(\xi^{2}_{{\rm D},i}+\frac{\xi^{2}_{\rm H}\xi^{2}_{{\rm G},i}}{L}\right)}\right)}{\Gamma(N-M+1)}.
\label{poutDirectRIS}
\end{align}

\subsection{Joint coherent/noncoherent detection}
Consider the reasonable scenario when the receiver knows the channel status of the direct link and knows nothing about the (instantaneous) RIS-enabled small-scale channel fading. The received signal here is identical to \eqref{receivedCSIdirect}. Yet, it is convenient for performance analysis reasons to follow an alternative approach by adopting the QR decomposition of the given channel fading matrix. Specifically, let $\mathbf{H}_{\rm D}\triangleq \mathbf{Q}\mathbf{R}$, where $\mathbf{Q}$ and $\mathbf{R}$ denote an $N\times N$ unitary matrix (with its columns representing the orthonormal ZF nulling vectors) and $N\times M$ upper triangular matrix, respectively. Then, $\mathbf{W}=\mathbf{Q}^{\mathcal{H}}$ and the received signal becomes
\begin{align}
\nonumber
\mathbf{r}&=\mathbf{Q}^{\mathcal{H}}\left[\sqrt{p}\left(\mathbf{H}_{\rm D}+\mathbf{H}\mathbf{\Phi}\mathbf{G}\right)\mathbf{s}+\mathbf{n}\right]\\
\nonumber
&=\sqrt{p} \left(\mathbf{R}+\mathbf{Q}^{\mathcal{H}}\mathbf{H}\mathbf{\Phi}\mathbf{G}\right)\mathbf{s}+\mathbf{Q}^{\mathcal{H}}\mathbf{n}\\
&\approx \sqrt{p} \left(\mathbf{R}+\mathbf{Q}^{\mathcal{H}}\mathbf{A}\mathbf{\Psi}^{1/2}\right)\mathbf{s}+\mathbf{Q}^{\mathcal{H}}\mathbf{n},
\label{receivedProposed}
\end{align}
where the last approximation implies that $L\gg 1$ by referring back to \eqref{CLTapprox}. By closely observing \eqref{receivedProposed}, the (unknown) RIS-enabled channel term $\mathbf{H}\mathbf{\Phi}\mathbf{G}\mathbf{s}$ carries transmit symbol information; thus, it can be treated as \emph{a signal rather than (colored) noise}. To this end, prior to symbol decoding, the post-detected signal in \eqref{receivedProposed} passes through a simple noncoherent detector (e.g., envelope detection) such that the received SNR of the $i^{\rm th}$ stream reads as
\begin{align}
\gamma_{i}\approx \frac{p \left|\mathbf{R}+\mathbf{Q}^{\mathcal{H}}\mathbf{A}\mathbf{\Psi}^{1/2}\right|^{2}_{i,i}}{\mathbf{Q}^{\mathcal{H}}\mathbb{E}\left[\mathbf{n}\mathbf{n}^{\mathcal{H}}\right]\mathbf{Q}}\overset{\text{d}}=p \left|\mathbf{R}+\mathbf{A}\mathbf{\Psi}^{1/2}\right|^{2}_{i,i}.
\label{receivedProposedSNR}
\end{align}
According to the unitary property applied on zero-mean Gaussian vectors \cite[Thm. 1.5.5]{b:multivariate}, i.e., $\mathbf{Q}^{\mathcal{H}}\mathbf{n}\overset{\text{d}}=\mathbf{n}$ and $\mathbf{Q}^{\mathcal{H}}\mathbf{A}\overset{\text{d}}=\mathbf{A}$, the derived SNR expression in \eqref{receivedProposedSNR} is accurate and reveals a clear physical meaning. Particularly, in most practical system setups, RIS-enabled channel links may be much more impactful than the direct transmitter-receiver ones (say, much shorter distances), which in turn pronounce the contribution of the $\mathbf{H}\mathbf{\Phi}\mathbf{G}\mathbf{s}$ signal term to the overall system performance more emphatically in comparison to the conventional approach.  

Given $\mathbf{H}_{\rm D}$ (and, thus, $\mathbf{R}$), the received SNR of the $i^{\rm th}$ stream as per \eqref{receivedProposedSNR} is distributed as $\gamma_{i}\overset{\text{d}}=\chi^{2}_{2}(p |r_{i,i}|^{2},\sigma^{2}_{i})$, where $r_{i,j}$ is the entry at the $i^{\rm th}$ row and $j^{\rm th}$ column of $\mathbf{R}$ as well as $\sigma^{2}_{i}\triangleq p \xi^{2}_{\rm H}\xi^{2}_{{\rm G},i}/(2 L)$. It follows that
\begin{align}
P_{\gamma_{i}|p |r_{i,i}|^{2}}(\gamma_{\rm th}|y)\approx 1-Q_{1}\left(\sqrt{\frac{y}{\sigma^{2}_{i}}},\sqrt{\frac{\gamma_{\rm th}}{\sigma^{2}_{i}}}\right),
\label{cdfconditional}
\end{align}
with $f_{p |r_{i,i}|^{2}}(y)=y^{N-M} \exp(-y/p \xi^{2}_{{\rm D},i})/[(N-M)!(p \xi^{2}_{{\rm D},i})^{N-M+1}]$. Finally, the unconditional outage probability of the $i^{\rm th}$ stream is derived by
\begin{align}
\nonumber
&P_{\gamma_{i}}(\gamma_{\rm th})\approx \int^{\infty}_{0}P_{\gamma_{i}||r_{i,i}|^{2}}(\gamma_{\rm th}|y) f_{|r_{i,i}|^{2}}(y) dy\\
\nonumber
&=1-\exp\left(-\frac{L \gamma_{\rm th}}{p \xi^{2}_{\rm H}\xi^{2}_{{\rm G},i}}\right)\Bigg[1+\sum^{N-M}_{l=0}\frac{L^{2} \gamma_{\rm th}\left(\frac{L \xi^{2}_{{\rm D},i}}{\xi^{2}_{\rm H}\xi^{2}_{{\rm G},i}}+1\right)^{-l}}{L p \xi^{2}_{\rm H}\xi^{2}_{{\rm G},i}+\frac{p \left(\xi^{2}_{\rm H}\xi^{2}_{{\rm G},i}\right)^{2}}{\xi^{2}_{{\rm D},i}}}\\
&\times {}_1F_{1}\left(l+1;2;\frac{L^{2} \gamma_{\rm th}}{L p \xi^{2}_{\rm H}\xi^{2}_{{\rm G},i}+\frac{p \left(\xi^{2}_{\rm H}\xi^{2}_{{\rm G},i}\right)^{2}}{\xi^{2}_{{\rm D},i}}}\right)\Bigg],
\label{poutunconditional}
\end{align}
where the latter result is obtained by utilizing \cite[Eq. (12)]{j:sofotasiosmarcum} and after some straightforward manipulations.\footnote{Typically, ${}_1F_{1}(\cdot)$ is an infinite series representation. Fortunately, in this certain case, the said function includes integer-valued parameters. Therefore, it can be reduced to a finite series representation with the aid of \cite[Eqs. (07.20.03.0026.01) and (07.20.03.0025.01)]{wolfram} including elementary-only functions. Doing so, \eqref{poutunconditional} denotes an efficient closed-form expression of the considered system outage performance.}

\subsection{Engineering Insights and Outcomes}
In practice, the considered approach can be realized in the uplink of a (distributed) MIMO system operating with the aid of an intermediate RIS. The independent Rayleigh fading assumption is valid whenever a dominant (line-of-sight) channel link is absent within a rich-scattering environment; e.g., in a dense outdoor urban terrestrial. The considered spatial multiplexing approach relies on the fact that $N\geq L\geq M$, which in various cases it holds that $N\geq L\gg M$ (for instance in massive MIMO deployments). For such a system setup, the channel is usually being estimated via pilot signaling prior to the data communication phase. Eventually, it turns out that the pilot length reaches to \cite{j:ZhangRui2020} $N L M+N M$ channel uses (i.e., time instances), which requires rather long pilot sequences; especially for quite large $\{N,L\}$ arrays. This process is time-consuming and may even do not comply with fundamental boundaries, such as the transmission block coherence time interval, which renders the problem quite challenging and prohibitive for practical applications \cite{j:WangLiu2020}.

On another front, in current work, we adopt the conventional channel status acquisition between (only) the direct transmitter/receiver faded links, while the RIS elements employ random rotations. We capitalize on the indirect impact caused by the channel gain of the RIS-enabled links in a noncoherent yet practically feasible basis. It is noteworthy that when adopting the considered approach, the length of pilot symbols is only $N M$, which is a viable solution. In addition, an outage/error floor is evident when partial CSI is available (either of the direct or RIS link). On the other hand, the received SNR grows unboundedly only when full CSI is available (ideal case, hard to implement) or when adopting the joint coherent/noncoherent approach. Next section cross-compares the performance of the above schemes and manifests the efficiency of the performance/complexity tradeoff provided by the considered semi-blind joint detection scheme.

\section{Numerical Results and Concluding Remarks}
In this section, the derived analytical results are verified via numerical validation, whereas they are cross-compared with corresponding Monte-Carlo simulations. Also, without loss of generality and for the sake of clarity, hereinafter, we assume an identical statistical profile for each transmitted stream (i.e., $\xi^{2}_{{\rm D},i}\triangleq \xi^{2}_{\rm D}\:\:\forall i$ and $\xi^{2}_{{\rm G},i}\triangleq \xi^{2}_{\rm G}\:\:\forall i$) in order to evaluate more concretely the overall system performance and obtain more impactful insights. In what follows, line-curves and solid dot-marks denote the analytical and simulation results, respectively. Labels `\emph{D}', `\emph{RIS}', `\emph{D+RIS}', and `\emph{Joint}' correspond, respectively, to the cases when CSI is known only for the direct link, RIS link, direct and RIS links, as well as the joint coherent/noncoherent approach with only the direct link known. Finally, we set $R=3$ bps/Hz. 

\begin{figure}[!t]
\centering
\includegraphics[trim=2.0cm .5cm 2.5cm 1.1cm, clip=true,totalheight=0.45\textheight]{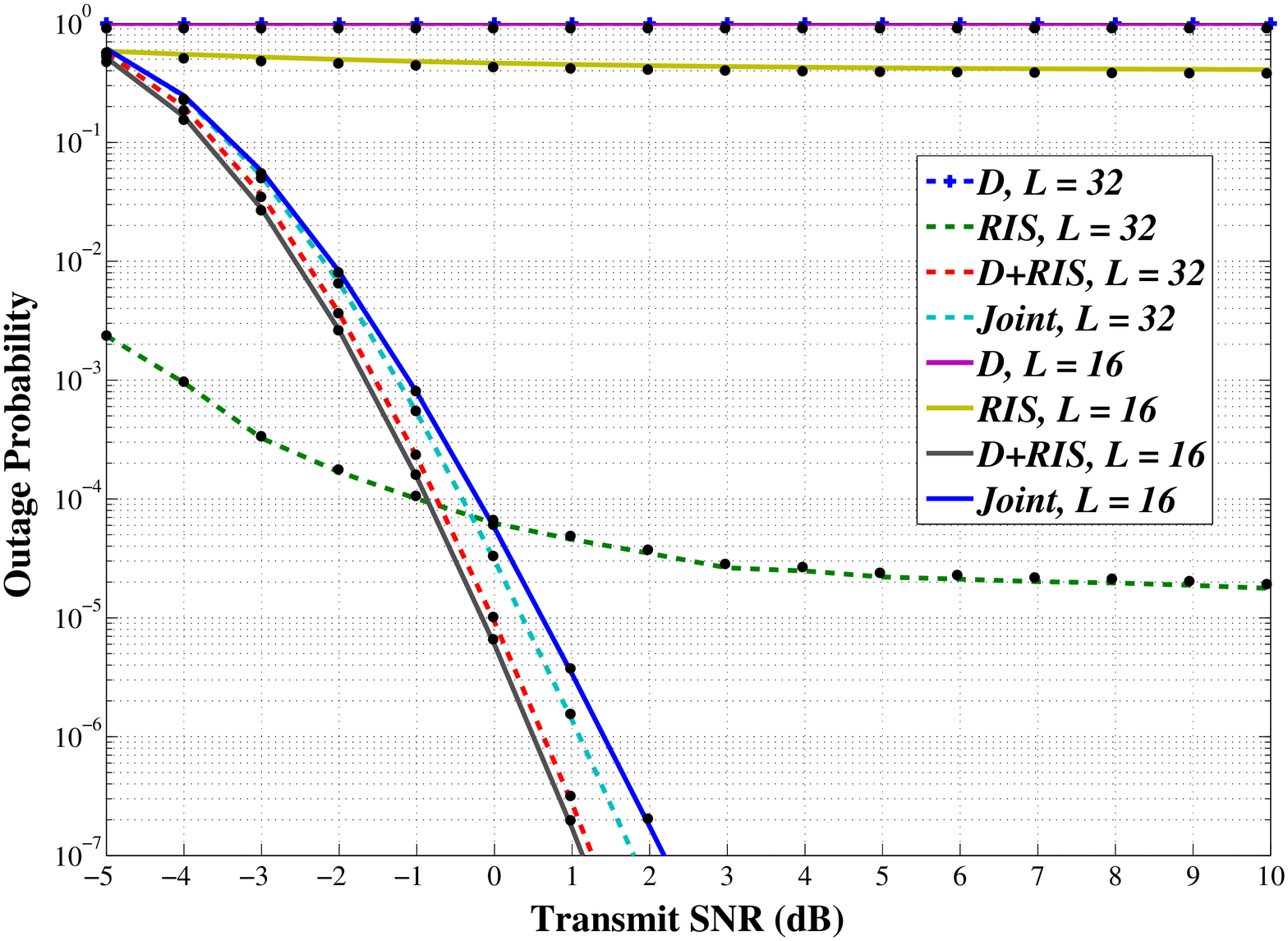}
\caption{Outage probability vs. various transmit SNR values and different RIS arrays, when $M=12$ and $N=32$. Also, it is assumed that $\{\xi^{2}_{\rm D},\xi^{2}_{\rm G},\xi^{2}_{\rm H}\}=1$.}
\label{fig1}
\end{figure}

In Fig.~\ref{fig1}, the system outage performance is illustrated vs. various transmit SNR regions with a particular emphasis on the relatively low SNR regime. The multiuser case of $M=12$ independent transmitted streams is modeled under large arrays of $N=32$ receive antennas and either $L=16$ or $32 (=N)$ RIS elements. Obviously, the detrimental effect of the partial CSI (present at the `\emph{D}' and `\emph{RIS}' cases) appears due to the unavoidable outage/error floor. Most importantly, the performance difference between the full-CSI (coherent detection) `\emph{D+RIS}' approach against to the noncoherent joint approach is quite small (less than $1$dB), while it is being reduced for an increasing number of RIS elements. 

\begin{figure}[!t]
\centering
\includegraphics[trim=2.0cm .5cm 2.5cm 1.1cm, clip=true,totalheight=0.45\textheight]{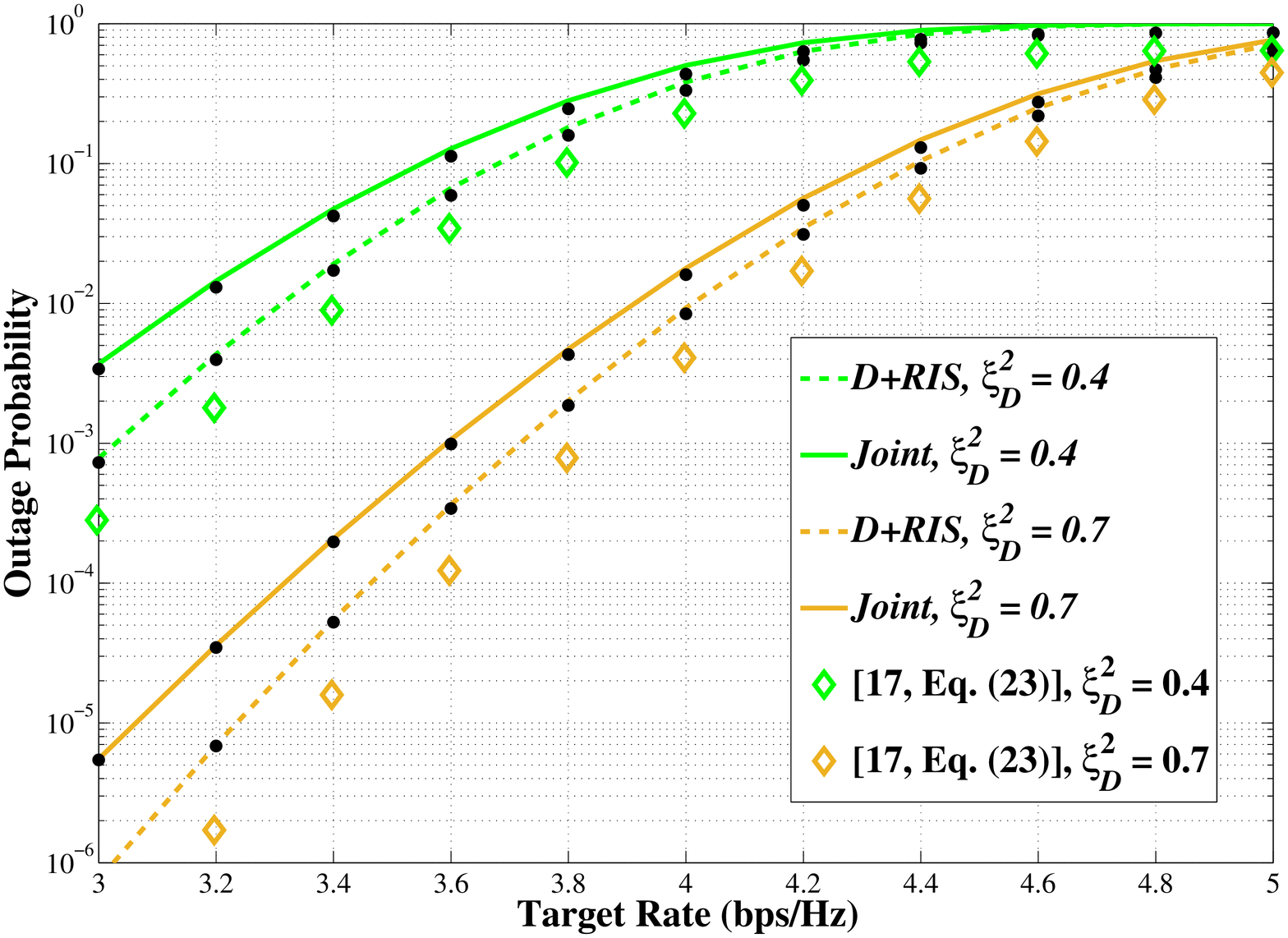}
\caption{Outage probability vs. different target data rates, where $N=32$, $M=14$ and $L=16$. Also, it is assumed that $\xi^{2}_{\rm H}=\xi^{2}_{\rm G}=0.7$, while $p=3$dB.}
\label{fig2}
\end{figure}

In Fig.~\ref{fig2}, the system outage performance is illustrated for various target data rates, when the large-scale channel gain of the direct link, $\xi^{2}_{\rm D}$, is equal or less to the individual RIS-enabled channel gains, correspondingly. This implies longer (at least equal) distances of the direct transceiver links in comparison to the individual RIS-enabled ones. As expected, the performance gets worse for an increased $R$ and reduced $\xi^{2}_{\rm D}$. It is verified again that the coherent scheme slightly outperforms the corresponding joint approach at the expense of a considerably higher computational overhead. {\color{black}Furthermore, the suboptimal approach under perfect CSI in \cite{j:ZhangRui2020} is presented for performance comparison reasons. In fact, \cite[Algorithm 1]{j:ZhangRui2020} introduces a joint optimization method of both power transmission and phase alignment at RIS. However, only the phase optimization is implemented in Fig.~\ref{fig2} (for a fair comparison to the proposed approach), as per \cite[Eq. (23)]{j:ZhangRui2020}. Notably, the suboptimal scheme in \cite{j:ZhangRui2020} outperforms the considered random-phase approach; yet, at the cost of an increased computational complexity and a considerably higher signaling overhead ($N L M+N M$ against only $N M$ provided by the proposed random-phase approach).}

Overall, the  outcomes derived from Figs.~\ref{fig1} and~\ref{fig2} are encouraging since the joint detection scheme comes with a significant reduction of the signaling (pilot) overhead which is in the order of $N L M$; whereby, it reflects a notable gain especially in massive MIMO deployments with large RIS arrays. Such a beneficial feature is achieved by introducing a manageable computational burden via a cost-efficient hardware gear for the noncoherent detection so as to capture the RIS-enabled channel gains.

\bibliographystyle{IEEEtran}
\bibliography{IEEEabrv,References}

\vfill

\end{document}